\documentclass{llncs}
\usepackage{times}
\usepackage{makeidx}  
\usepackage{graphicx}
\usepackage{amssymb}
\usepackage{latexsym}
\usepackage{amsmath}
\usepackage{float}
\usepackage{wrapfig}
\usepackage{subfigure}
\usepackage{graphics}
\begin{document}
\newtheorem{axiom}{Axiom} 
\newtheorem{arule}{Rule} 
\newtheorem{algorithm}{Algorithm} 
 \newcommand{\Rtrademark}{$^{\text{\textcircled{\tiny R}}}$}
\title{Fast and Generalized Polynomial Time Memory Consistency Verification}
\author{Amitabha Roy, Stephan Zeisset, Charles J. Fleckenstein, John C. Huang}
\institute{Intel Corporation\\
\email\{amitabha.roy,stephan.zeisset,chuck.fleckenstein\}@intel.com,jhuangtw@umich.edu}
\maketitle
\bibliographystyle{unsrt}
\begin{abstract}
The problem of verifying multi-threaded execution against the memory consistency model of 
a processor is known to be an NP hard problem. However polynomial time algorithms
exist that detect almost all failures in such execution. These are often used in practice for
microprocessor verification. We present a low complexity and fully parallelized algorithm
to check program execution against the processor consistency model. In addition our algorithm
is general enough to support a number of consistency models without any degradation in performance.
An implementation of this algorithm is currently used in practice to verify processors in the 
post silicon stage for multiple architectures.
\end{abstract}

\section{Introduction}
Verifying processor execution against its stated memory consistency model is an important problem in both 
design and silicon system verification. Verification teams for a microprocessor are often concerned 
with the memory consistency model visible to external customers such as system programmers.
In the context of multi-threading, both in terms of Simultaneous Multi Threading(SMT) and 
Chip Multi Processing(CMP), Intel\Rtrademark \footnote{Intel\Rtrademark is a trademark or registered trademark
of Intel Corporation or its subsidiaries in the United States and other countries.} and other CPU manufacturers are
increasingly building complex 
processors and SMP platforms with a large number of execution threads. In this environment the memory consistency
model of microprocessors will come under close scrutiny, particularly by developers of multi-threaded 
applications and operating systems. Allowing any errors in implementing the consistency model to show
up as customer visible is thus unacceptable.
The problem we are concerned with is that of matching the result of executing a random set of load store 
memory operations distributed across processors, on a set of shared locations, against a memory 
consistency model. The algorithm should flag an error if the consistency model does not allow the observed execution results.
This forms the basis for Random Instruction Test(RIT) generators such as TSOTOOL$^*$\footnote{$^*$ Other names and brands
may be claimed as the property of others} \cite{tsotool} and Intel's
Multi Processor(MP) RIT environment. The Intel MP RIT Tool incorporates the algorithm in this paper.  
Formally, we concentrate on variations of the VSC (Verifying Sequential Consistency) problem \cite{vsc}. 
The VSC problem is exactly the problem described above, when restricted to sequential consistency. 
The general VSC problem is NP complete \cite{vscisnp}. The general coherence problem has also been shown 
to be NP complete \cite{cantin-spaa03-and-ms}. A formulation of VSC for more general memory 
consistency models was done in \cite{tsotool} 
where a polynomial time algorithm was presented for verifying a memory consistency model at the cost of correctness, 
although the incorrect executions missed were shown to be insignificant for the purpose of CPU verification. That work focused
almost exclusively on the Total Store Order(TSO) memory consistency model and presented a worst case $O(n^5)$ algorithm.
In this work, we present an efficient implementation of the basic algorithm in \cite{tsotool}. 
Our key contribution is to reduce the worst case complexity to $O(n^4)$ for \emph{any} memory consistency model using 
$\Theta(n^2)$ space. Although the work in \cite{tsofast} has reduced the complexity to $O(kn^3)$ 
where k is the number of processors, that algorithm assumes the TSO memory consistency model and does not generalize to other models. 
Our motivation for generalizing and improving it is Intel's complex verification environment, where microprocessors support as many as 
five different consistency models at the same time. The primary objectives of our algorithm design are simplicity, performance and 
seamless extendibility in the implementation to any processor environment, including the Itanium\Rtrademark\footnote{Itanium\Rtrademark 
is a trademark or registered trademark
of Intel Corporation or its subsidiaries in the United States and other countries.}. Another goal is enhanced 
support for debugging reported failures, which is crucial to reducing time to market for complex multi processors.

The algorithm we have developed is currently implemented in Intel's in house random test 
generator and is used by both the IA-32
and Itanium verification teams. We also present scalability results and 
a processor bug that was caught by the tool using this algorithm.
\section{Memory Consistency}
Consider a set of processors each of which executes a stream of loads and stores. These are done to a set of locations
shared across the processors. We are concerned with a global ordering of all the
loads and stores, which when executed serially leads to the same result. The strictest consistency model is the sequential consistency (SC)
model which insists that the only valid orderings are those that do not relax per processor program order between the memory operations.
 Relaxing restrictions between operations such as stores and loads leads to progressively weaker models
such as Total Store Order (TSO) and Release Consistency (RC). All these are surveyed in \cite{tutorial}.
We point out that in these orderings we refer to load executions 
and store executions. A load is considered performed(or executed) if no subsequent store to that location(on any processor) can change the 
load return value. A store is considered performed(or executed) if any subsequent load to that location (on any processor) returns its value. These are 
definitions from \cite{survey}. Any instruction 
on a modern pipelined processor has a number of phases and some, such as instruction fetch and retirement, occur in strict program 
order without regard to the memory consistency model. We are concerned only with ordering the load and store execution phases for 
instructions referring to memory.

\subsection{Formalism}
The terminology used in this paper is similar to \cite{tsotool}.
We use $;$ to denote program order and $\leq$ to denote global order. Thus $A;B$ and $A\leq B$ mean that B follows A in 
program order and global order respectively. The fundamental operations in our test consist of 
$L_a^i$ and $S_a^i$ which are
loads and stores respectively to location $a$ by processor $i$. We also consider $[L_a^i;S_a^i]$ which is an atomic load store operation.
Examples are XCHG in IA-32 \cite{isa2} and FETCHADD in Itanium \cite{itaniumisa}. We use $val(L_a^i)$ to denote the load return value of a 
load operation and $val(S_a^i)$ to denote the value stored by a store operation.

For any location $a$ we define the type of a location to be\\ $\mathit{Type}(a) \in \{WB, WT, WP, UC, WC\}$. The type of a location is 
the \emph{memory} type 
of the location. IA-32 \cite{ia32} supports all five memory types, Write Back (WB), Write Through (WT), Write Protect (WP), Write Combining(WC)
and Uncacheable. Itanium \cite{itanium} supports only three, WB, WC and UC. In addition to cacheability and 
write through implications of these memory types, they also affect the consistency model. 
\subsection{Axioms and Orders}
Both $\leq$ and $;$ are transitive, reflexive and antisymmetric orders. The program order is limited to operations on the same processor
while the global order covers all operations across all processors. We also define $A < B$ to mean $A \leq B$ and $A \neq B$.

We define the following axiom to support atomic operations.
\begin{axiom}[Atomic Operations\\]
\label{axm:atomic}
$[L_a^i;S_a^i]\Rightarrow (L_a^i \leq S_a^i) \bigwedge (\forall S_b^j: (S_b^j \leq L_a^i) \bigvee (S_a^i \leq S_b^j))$
\end{axiom}  
As a result of this, we can treat atomic operations as a single operation for verification.
We assume the following two axioms to hold, the bare minimum to be able to use the basic algorithm proposed in \cite{tsotool}.
\begin{axiom}[Value Coherence\\]
\label{axm:wratomic}
$val[L_a^i] \in \{ val[\stackrel{\mathit{Max}}{\leq} {S_a^k | S_a^k < L_a^i}] , val[\stackrel{\mathit{Max}}{;} {S_a^i | S_a^i;L_a^i}] \}$  
\end{axiom}
The value returned by a read is from either the most recent store in program order or the most recent store in global order.
This is intuitive for a cache coherent system. Note that the most recent store in program order may not be a preceding store in global
order. This is because many architectures including Intel ones can support the notion of store forwarding, which allows a store to be forwarded
to local loads before it is made globally visible. Also, in the test a load may occur before any store to that location in which case 
it returns the
initial value of that location. Such cases are handled by assuming a preliminary set of stores that write initial values to locations. 
The store values to a location and initial value of the location are chosen to be unique by the test generator.
This allows the axiom to be applied after the test is completed to link a load to the store that it reads.

\begin{axiom}[Total Store Order\\]
\label{axm:tso}
$\forall S_a^i, S_b^j ((S_a^i \leq S_b^j) \bigvee (S_b^j \leq S_a^i))$.
\end{axiom}
Unlike \cite{tsotool}, we have avoided imposing any additional constraints between operations on the same processor. Rather, we allow
these constraints to be dynamically specified. This allows us to parameterize the same algorithm to work across CPU architectures
(Itanium and IA-32) and processor generations (Intel NetBurst\Rtrademark\footnote{Intel NetBurst\Rtrademark is a trademark or registered trademark
of Intel Corporation or its subsidiaries in the United States and other countries.} and P6 in the case of IA-32).

Define $\mathit{Ops} = \{L, S, X\} $ to be the allowed types of an operation. Thus we can define $\mathit{Type}(L_a^i)=L$, $\mathit{Type}(S_a^i)=S$ 
and 
$\mathit{Type}([L_a^i;S_a^i])=X$. We also define $\mathit{Loc}(Op)$ to return the memory location used by the operation. For example
$\mathit{Loc}(L_a^i) = a$.

We can then define the constraint function\\ $f: (\mathit{Ops} X \{WB, WP, WT, WC, UC\})^2 \rightarrow \{0, 1\}$.
This is used to impose the dynamic set of constraints:
\begin{definition}[Local Ordering]
\label{def:local}
[$O_1 ; O_2 \mathit{and}$\\ $f((\mathit{Type}(O_1), \mathit{Type}(\mathit{Loc}(O_1)), (\mathit{Type}(O_2),\mathit{Type}(\mathit{Loc}(O_2)))) =1] \Rightarrow O_1 \leq O_2$\\
If the LHS of the implication is satisfied we call $O_1$ and $O_2$ as locally ordered memory operations.
\end{definition}

As an example, from \cite{ia32} we know that Write back stores do not bypass each other. Hence f((S, WB),(S,WB))=1. However,
write combining stores are allowed to bypass each other and hence f((S, WC), (S,WC))=0. There are other more subtle orderings which vary between processor generations and in this case we obtain appropriate ordering functions from the CPU architects or designers.
\section{Algorithm}
\label{sec:algo}
Our objective is an algorithm that takes in the result of an execution and flags violation of the memory consistency model.
The basic algorithm in \cite{tsotool} that we extend uses constraint graphs to model the execution. There have been similar 
approaches in the past too, such as \cite{cgraph} and an approach to the same problem using constraint solvers \cite{ganesh}.

We model the execution as a directed graph G=(V, E) where the nodes represent memory operations and the edges represent 
the $\leq$ global order. However, as in \cite{tsotool}, we do not put self edges although the relation is reflexive.
Thus if $O_1 \leq O_2$ then we add an edge from the node for $O_1$ to that for $O_2$. For brevity, we refer to operations and their 
corresponding nodes by the same name. $A \rightarrow B$ means there is an edge from $A$ to $B$ while 
$A \rightarrow_P B$ means there is a path from $A$ to $B$.

Based on the per processor ordering imposed by our ordering function $f$, we can immediately add static edges to the graph.

\begin{arule}[Static Edges]
\label{rl:static}
For every pair of nodes $O_1$ and $O_2$ such that they are locally ordered by definition 
\ref{def:local}, add the edge $O_1 \rightarrow O_2$.
\end{arule}

After execution of the test, we determine a function $Reads$ in a preprocessing step (operating on loads) 
such that $Reads(L_a^i)=S_a^j$ if $L_a^i$ reads $S_a^j$.
Otherwise (the case where the initial value for the location is read), $Reads(L_a^i)=Sentinel$, a special sentinel node.
We add edges from $Sentinel$ to all other store nodes in the graph. This is the
same construction as described in \cite{tsotool}. From the value axiom we know that any read that returns the value of 
a remote write must have occurred after the
remote write has been globally observed. This allows us to add observed edges to the graph based on the values 
returned by the loads in the test. Note that for the rules below we treat an atomic operation as both a load
and a store.

\begin{arule}[Observed Edge]
\label{rl:obs}
For every load $L_a^i$, if $Reads(L_a^i)=S_a^j$ where $i \neq j$, or if $Reads(L_a^i)=Sentinel$,
add the edge  $Reads(L_a^i)\rightarrow L_a^i$.
Note that since stores to the same location write unique values and all locations are initialized to hold
unique values, value equivalence means that the load must have read that store.
\end{arule}
The next few set of edges are essentially inferred from the value axiom. Hence they are called inferred edges.
\begin{arule}[Inferred Edge 1]
\label{rl:inf1}
If $Reads(L_a^i)= S_a^j$ and $i \neq j$ then for every $S_a^i$ such that $S_a^i;L_a^i$ add the edge
$S_a^i \rightarrow S_a^j$. This follows from the value axiom since the alternative global order would mean the
load should read the local store.
\end{arule}
\begin{arule}[Inferred Edge 2]
\label{rl:inf2}
If $Reads(L_a^i)=S_a^j$ then for every $S_a^k$ such that $S_a^k \rightarrow_P L_a^i$ and $S_a^k \neq S_a^j$, 
add the edge $S_a^k \rightarrow S_a^j$. This follows from the value axiom since the alternative global order would mean 
that the load should read $S_a^k$.
\end{arule}
\begin{arule}[Inferred Edge 3]
\label{rl:inf3}
If $Reads(L_a^i)=S_a^j$ then for every $S_a^k$ such that $S_a^j \rightarrow_P S_a^k$ add the edge
$L_a^i \rightarrow S_a^k$. This follows from the value axiom since the alternative global order would mean 
that the load should read $S_a^k$.
\end{arule}
\subsection{Basic Algorithm}
The basic algorithm described in \cite{tsotool} can now be summarized as follows:
{\small
\begin{enumerate}
\item \label{ba:pre}Compute the $Reads$ function in a preprocessing step.
\item \label{ba:static}Apply rule \ref{rl:static} to add all possible edges.
\item \label{ba:obs}Apply rule \ref{rl:obs} to add all possible edges.
\item \label{ba:inf}Apply rules \ref{rl:inf1}, \ref{rl:inf2} and \ref{rl:inf3}.
\item \label{ba:fix}If any edges were added in step \ref{ba:inf} go back to step \ref{ba:inf} else go to step \ref{ba:check}
\item \label{ba:check}Check the graph for cycles. If any are found, flag an error.
\end{enumerate}
}
\begin{wrapfigure}{L}[0in]{5cm}
\subfigure{
\framebox{\includegraphics[trim=2cm 6cm 4cm 6cm, clip, height=4cm, angle=270, width=4cm]{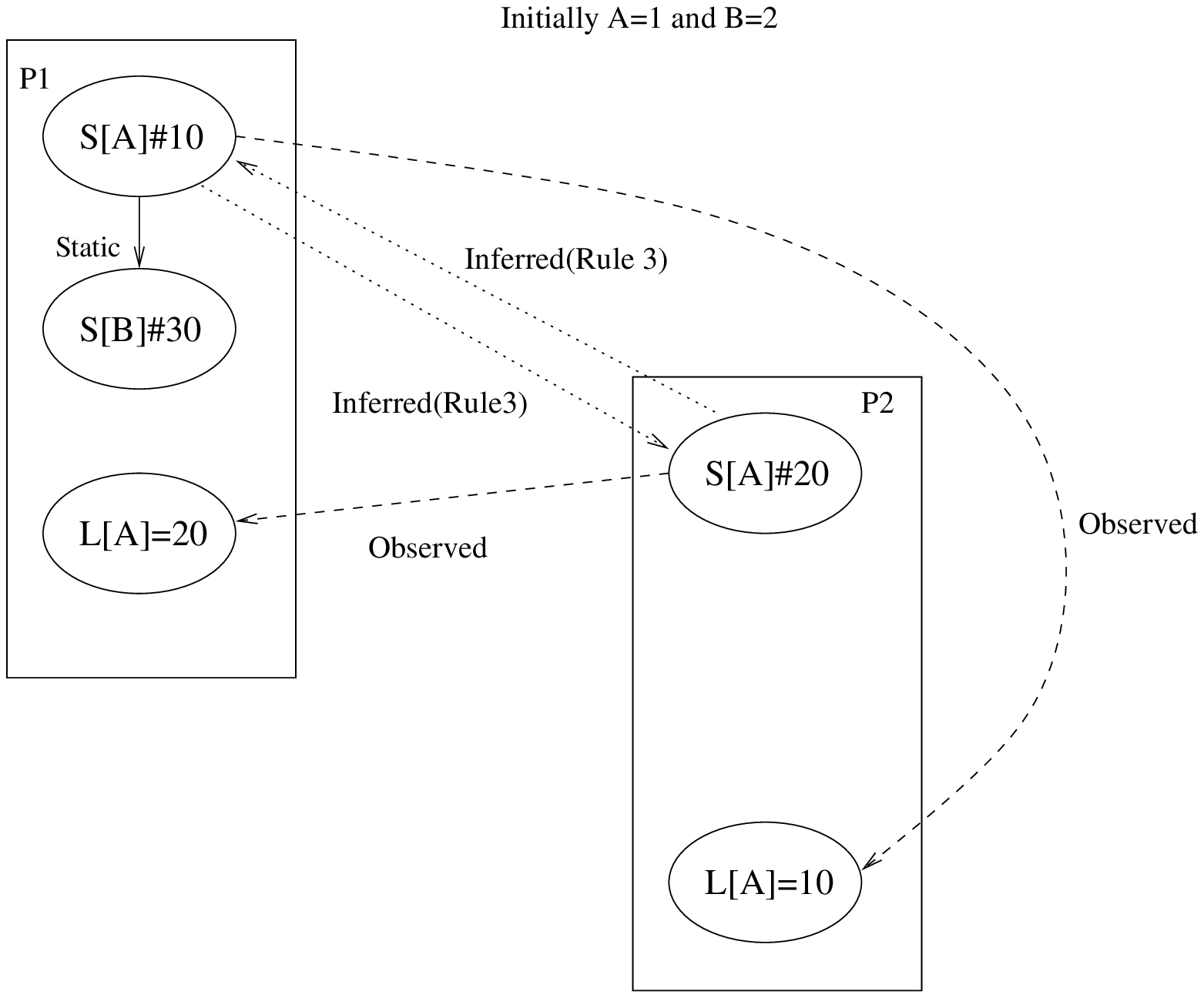}}}
\caption{Example of an incorrect execution with graph edges added}
\label{fig:example1}
\end{wrapfigure}
An example of this algorithm applied to an execution is shown in Figure \ref{fig:example1}. 
We use the notation $S[X]\#V$ for write $V$ to location $X$, and $L[X]=V$ for read from location 
$X$ returns value $V$.

Computing the $Reads$ function is $O(n^2)$ since we need to examine all pairs of loads and stores.
Steps \ref{ba:static} and \ref{ba:obs} are of cost $O(n^2)$ since we examine all pairs of nodes. 
Step \ref{ba:inf} involves determining the relationship
$A \rightarrow_P B$ for $O(n)$ nodes. This costs $O(n^2)$ for each node (assuming a depth first 
search, as one of the obvious options) and hence $O(n^3)$ overall. Since the fixed
point iteration imposed by steps \ref{ba:inf} and \ref{ba:fix} may loop for at most $O(n^2)$ adding one edge on each iteration, we
have a worst case complexity of $O(n^5)$. The detailed analysis is in \cite{tsotool}. There has been 
a subsequent improvement published in \cite{tsofast} that reduces the complexity to $O(kn^3)$. Its correctness 
requires that there are a constant number of ordered lists on each processor. This is true because all loads
and all stores are ordered on a processor in the TSO consistency model that they have considered. Unfortunately this
does not hold true for both the IA-32 \cite{ia32} and Itanium \cite{itaniumspec} memory models for various 
memory types (consider WC stores). Hence the formulation in \cite{tsofast} is not general enough.
\subsection{Graph Closure}
The primary contributor to the $O(n^5)$ complexity is deciding whether $A \rightarrow_P B$ holds.
All other operations can be efficiently implemented and do not seem to hold any opportunity for improvement, given our
goal of generality. Hence, we decided to focus on the problem of efficiently determining $A \rightarrow_P B$. A solution is
to compute the transitive closure of the graph. We first label all the nodes in the directed graph under consideration, 
$G=(V, E)$ by natural numbers using the bijective mapping function $g:V \rightarrow \{1 .. n\}$ where $\mid V\mid=n$. 
We can then represent $E$ by the familiar $n$ square adjacency matrix $A$ such that $(U,V) \in E \Leftrightarrow A[g(U),g(V)]=1$.

For transitive closure of the graph we seek the closed form of the adjacency matrix $A$ such that 
$U \rightarrow_P V \Leftrightarrow A[g(U),g(V)]=1$. A well known algorithm for computing the transitive 
closure of a binary adjacency matrix is Warshall's algorithm\cite{warshall}. 
Before giving Warshall's algorithm, we first define some convenient notation and
functions to transform the connectivity matrix. $AddEdge(x, y)$ stands for : set $A[x,y]=1$. $Subsume(x,y)$ 
is defined as $\forall z$ such that$A[y,z]=1$, $AddEdge(x, z)$. The subsume function causes all neighbors of node 
$g^{-1}(y)$ to also become neighbors of node $g^{-1}(x)$ in the adjacency matrix representation.\\\\
\fbox{
\begin{minipage}{2in}
\setlength{\parindent}{2mm}
{\small
\textbf{\underline{Warshall's Algorithm:}}\\
for all $j \in \{1 .. N\}$\\
\indent for all  $i \in \{1 .. N\}$\\
\indent \indent if($A[i,j]=1$)\\
\indent \indent \indent $Subsume(i,j)$\\
\indent \indent end if\\
\indent end for\\
end for\\
}
\end{minipage}
}    
\fbox{
\begin{minipage}{2.2in}
\setlength{\parindent}{2mm}
{\small
\textbf{\underline{Incremental Warshall's Algorithm:}}\\
for all $j \in \{1 .. N\}$\\
\indent for all  $i \in \{1 .. N\}$\\
\indent \indent if($A[i,j]=1$ and\\
\indent \indent ($Changed[j]=1$ or $Changed[i]=1$))\\
\indent \indent \indent $Subsume(i,j)$\\
\indent \indent end if\\
\indent end for\\
end for}
\end{minipage}}\\\\
\textbf{Incremental Graph Closure:}
Although Warshall's algorithm will compute the closed form of the adjacency matrix, any edge added by $AddEdge$ will cause the
matrix to lose this property since new paths may be available through the added edge. Hence we need an algorithm which when 
given a closed adjacency matrix and some edges added \emph{efficiently} recomputes the closure.

We assume that when adding edges to node $U$, we mark that node as changed by setting the corresponding 
bit in the change vector $Changed[g(U)]=1$. We can now rerun Warshall's algorithm \emph{restricted} to only those nodes
which have either changed themselves, or are connected in the current adjacency matrix to a changed node. This is shown in 
pseudo-code as incremental Warshall's algorithm.\\

\textbf{Correctness:}
The restricted Warshall's algorithm clearly terminates. 
Now, consider any new path as a result of addition of edges to the graph,\\
$(U_1,U_2), (U_2,U_3), ... ,(U_{m-1}, U_m)$. There is at least one edge $(U_i,U_{i+1})$ such
that $Changed[i]=1$. We need to show that $A[g(U_1),g(U_m)]=1$ at termination.
Since the matrix was already closed, we can eliminate sub-paths consisting only of edges from the original 
graph. The endpoints of these sub-paths would be connected in $A$. 
Thus we can form a subset of the nodes on 
this path (in the same order) $V_1, V_2, V_2, V_3, ..., V_l$ where $\forall V_i$ 
($ i > 1$) either $Changed[g(V_i)]=1$ or $Changed[g(V_{i-1})]=1$. Also, $\forall i$, $A[g(V_i),g(V_{i+1})]=1$ and 
we have $U_1 = V_1$ and $U_m=V_m$.

Observe that for every $V_i$ if $Changed[g(V_i)]=1$ then $Subsume(g(V_i), g(V_{i+1}))$ is called.
Otherwise if this is not the last node in the path, $Changed[g(V_{i+1})]=1$ and
$A[g(V_i),g(V_{i+1})]=1$. Hence, $Subsume(g(V_i), g(V_{i+1}))$ will always be called.

Using this observation, we can argue that we run Warshall's algorithm on a subgraph consisting
only of the path $V_1,V_2,V_2,V_3, ...,V_l$ (since those are connected in the adjacency matrix).
As Warshall's algorithm is correct \cite{warshall} we can conclude\\ $A[g(V_1),g(V_l)]=1$ at termination.
Since $V_1=U_1$ and $V_l=U_m$ by construction, we have $A[g(U_1),g(U_m)]=1$.

It is trivial to show that the incremental update adds no incorrect edges, since 
$A[i,j]=1$ is a precondition to the $Subsume(i,j)$.

\textbf{Complexity:}
An important observation is that the complexity of the incremental update is
$O(mn^2)$ where the number of changed nodes is $O(m)$. This is because the
subsume step takes $O(n)$ and for each node, $Subsume$ can only
be called at worst $O(m)$ times, if it is connected to all the changed nodes.
At worst all $O(n)$ nodes satisfy the precondition for subsume and hence the $O(mn^2)$
complexity.

\subsection{Final Algorithm:}
\label{sec:fa}
We describe algorithms to implement the rules for adding
observed and inferred edges in Table \ref{Tab:edge}. Recall that our graph is G=(V, E) and
the vertices correspond to memory operations in the test.Also, 
for ease of specification we have allowed atomic read modify write operations to be 
treated as both stores $Type(Op)=S$ and loads $Type(Op)=L$.
\begin{table}
\setlength{\parindent}{2mm}
\begin{small}
\begin{tabular}{l}
\\\hline
\textbf{Algorithm for adding edges:}\\\hline
\textbf{\underline{Static Edges:}}\\
 for all $O_1 \in V$\\
 \indent for all $O_2 \in V$ such that $O_1 \neq O_2$\\
 \indent \indent If $O_1$ is locally ordered after $O_2$ as per definition \ref{def:local}then $AddEdge(g(O_2),g(O_1))$\\
 \indent end for\\
 end for\\
 \textbf{\underline{Observed Edges:}}\\
 for all $O_1 \in V$ such that $type(O_1)=L$\\
 \indent for all $O_2 \in V$ such that $type(O_2)=S$\\ 
 \indent \indent If $val(O_1)=val(O_2)$\\
 \indent \indent \indent set $Reads(O_1)=O_2$\\
 \indent \indent \indent If $O_2$ is on a different CPU from $O_1$ then $AddEdge(g(O_2), g(O_1))$\\
 \indent \indent end If\\
 \indent end for\\
 \indent If no corresponding store is found for this load then $AddEdge(g(Sentinel), g(O_1))$\\
 \indent and set $Reads(O_1)=Sentinel$\\
 end for\\
 \textbf{\underline{Inferred Edge 1:}}\\
 for all $O_1 \in V$ such that $type(O_1)=L$\\
 \indent for all $O_2 \in V$ such that $type(O_2)=S$ and $O_2;O_1$ and $O_2 \neq Reads(O_1)$\\
 \indent \indent If $O_2$ is on a different CPU from $O_1$ then $AddEdge(g(O_2), g(Reads(O_1)))$ and set $Changed[g(O_2)]=1$\\
 \indent end for\\
 end for\\
 \textbf{\underline{Inferred Edge 2:}}\\
 for all $O_1 \in V$ such that $type(O_1)=L$\\
 \indent for all $O_2 \in V$ such that $type(O_2)=S$ and $A[g(O_2),g(O_1)]=1$\\
 \indent and $O_2 \neq Reads(O_1)$\\
\indent \indent $AddEdge(g(O_2), g(Reads(O_1)))$ and set $Changed[g(O_2)]=1$\\
\indent end for\\
end for\\
\textbf{\underline{Inferred Edge 3:}}\\
for all $O_1 \in V$ such that $type(O_1)=S$\\
\indent for all $O_2 \in V$ such that $type(O_2)=L$ and $A[g(Reads(O_2)),g(O_1)]=1$ \\
\indent \indent $AddEdge(g(O_2), g(O_1))$ and set $Changed[g(O_2)]=1$\\
\indent end for\\
end for
\\\hline
\end{tabular}
\end{small}
\caption{Pseudcode of Algorithm for Adding Edges}
\label{Tab:edge}
\end{table}
The ordering of for loops is not arbitrary as it may appear but rather has been carefully chosen to aid in parallelization 
as we demonstrate in section \ref{sec:parallel}. 

We now state the final algorithm used to verify the execution results.A benefit of our 
approach is that checking the graph for cycles is simply checking
whether $\exists i$ $A[i,i]=1$ since a cycle results in a self loop due to 
the closure. Additionally, note that we have merged the preprocessing step that links loads
to the stores they read, into the step to compute observed edges.
{\small
\begin{enumerate}
\item \label{fi:static}Apply rule \ref{rl:static} to add all possible edges.
\item \label{fi:obs}Apply rule \ref{rl:obs} to add all possible edges.
\item \label{fi:war}Apply Warshall's algorithm to obtain the closed adjacency matrix.
\item \label{fi:inf}Apply rules \ref{rl:inf1}, \ref{rl:inf2} and \ref{rl:inf3}.
\item \label{fi:fix}If any edges were added in step \ref{fi:inf} go to step 6 else go to step 8.
\item \label{fi:inc}Apply the incremental Warshall's algorithm to recompute closure and reset the changed vector.
\item \label{fi:loop}Go to step \ref{fi:inf}.
\item \label{fi:check}Check the graph for cycles. If any are found, flag an error.
\end{enumerate}}
\subsection{Complexity}
\begin{wrapfigure}{R}{5cm}
\subfigure{\framebox{\includegraphics[trim=2.85cm 4cm 2cm 5cm, clip, height=4cm, angle=270, width=4cm]{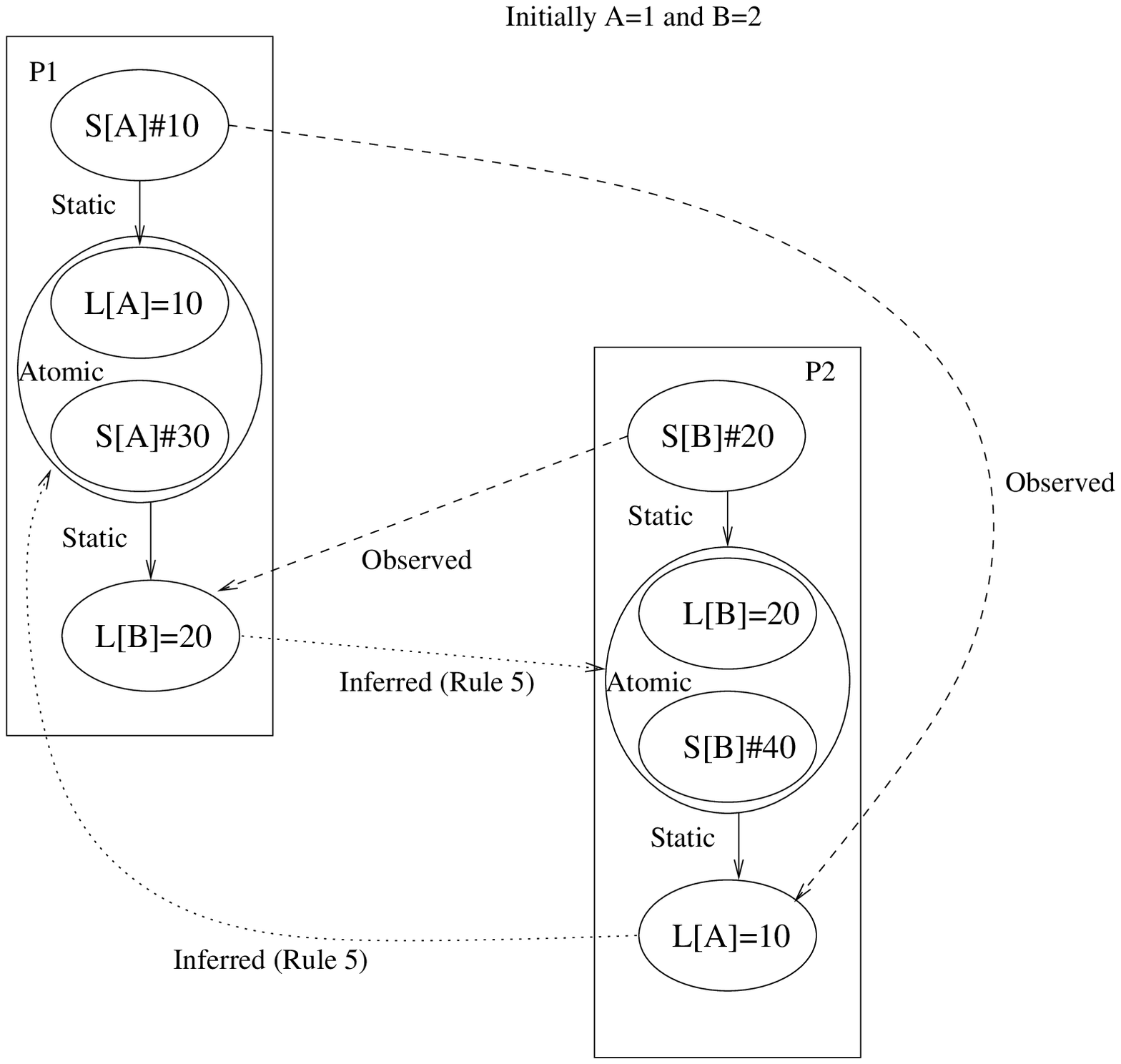}}}
\caption{Example of an actual processor bug}
\label{fig:example2}
\end{wrapfigure}
The analysis of complexity is straightforward.
Each of steps \ref{fi:static} and \ref{fi:obs} take $O(n^2)$ since they examine all pairs of nodes. Step \ref{fi:war} takes $O(n^3)$ 
as is shown in \cite{warshall}.
Each iteration of Step \ref{fi:inf} again takes $O(n^2)$ because we examine all pairs of nodes. Note that checking $A \rightarrow_P B$ is
now $O(1)$ thanks to the closed adjacency matrix. There are at most $O(n^2)$ edges to be added
and hence the worst case complexity for Step \ref{fi:inf} is $O(n^4)$. The remaining analysis is 
step \ref{fi:inc}. For this we note that the complexity is also $O(mn^2)$ when considered over \emph{all}
invocations. Since $m=O(n^2)$ (bounded above by the number of edges we can possibly add and thereby change nodes), we have $O(n^4)$ 
as the worst case complexity for step \ref{fi:inc}. Cycle checking in step \ref{fi:check} is simply $O(n)$ due to the closed form of 
the adjacency matrix.  Thus the overall complexity is $O(n^4)$ which meets our stated goal. Our overall space requirements are clearly
 $\Theta(n^2)$ due to the adjacency matrix.
\begin{wrapfigure}[30]{L}[0in]{7cm}
\fbox{
\begin{minipage}{2.5in}
\setlength{\parindent}{2mm}
{\small
\textbf{\underline{Algorithm PrintSomeCycle:}}\\
PossibleStart=$\{g^{-1}(i) \mid A[i,i]=1\}$\\
while PossibleStart is not empty\\
\indent StartNode=any node in PossibleStart\\
\indent PossibleStart=PossibleStart -\{StartNode\}\\
\indent CurrentList=$\{g^{-1}(i) \mid A[i,i]=1\}$ - StartNode\\
\indent GetCycleEdge(startNode,startNode)\\
end while\\\\
\textbf{\underline{Function GetCycleEdge:}}\\
GetCycleEdge(node Start, node Current)\\
If Algorithm(Current, Start) returns true\\
\indent print edge (Current, Start)\\
\indent PossibleStart=PossibleStart -\{Current\}\\
\indent return true\\
end If\\
for each node nextNode in CurrentList\\
\indent If Algorithm(Current, nextNode) returns true\\
\indent \indent CurrentList = CurrentList - \{nextNode\}\\
\indent \indent If GetCycleEdge(Start, nextNode) returns true\\
\indent \indent \indent	print edge (Current, nextNode)\\
\indent \indent \indent PossibleStart=PossibleStart -\{Current\}\\
\indent \indent \indent	return true\\
\indent \indent end If\\
\indent end If\\
end for\\
return false}
\end{minipage}}
\caption{Debug Algorithm}
\label{fig:debug}
\end{wrapfigure}
\section{Parallelization}
\label{sec:parallel}
One of the ways to mitigate the expense of an $O(n^4)$ algorithm is parallelization. With a test size of hundreds of 
memory operation per CPU, result validation time can easily overwhelm the verification process. For example 
consider a 4 way SMP platform
with hyperthreaded processors with a total of 8 threads and hence 800 operations. The way we have arranged the algorithm and
data structures allows us to easily do loop parallelization \cite{loop}.

The phases of the algorithm are Warshall's algorithm, incremental graph closure and the rule algorithms given in section \ref{sec:fa}.
The key observation is that in each case we always have no more than two nested for loops and there are no data dependences 
between iterations of the inner loop. The latter is true because no two iterations change the same node in the graph and hence
never write to the same element in the adjacency matrix.
We are not worried about considering edges added in previous iterations of the inner for loop of step \ref{fi:inf} (of the algorithm 
in \ref{sec:fa})
because such edges are considered in subsequent iterations, since we iterate to a fix point. Also the same element in the $Changed$ 
vector is not accessed
by two different inner loop iterations. Hence we can parallelize by distributing different iterations of the inner
for loop in each step across processors.
Since each inner for loop iterates over all nodes in the graph, this leads to a convenient data partitioning. We allocate each 
CPU running the
verification algorithm a disjoint subset of nodes in the graph. Each CPU executes the inner for loop in each phase only on nodes 
that it owns.
Note that each CPU still needs to synchronize with all other CPUs after completion of the inner for loop in each case (this is 
similar to the INDEPENDENT FORALL construct in High Performance Fortran).

\section{Implementation}
Intel's verification environment spans both architecture validation (Pre Silicon on RTL models) 
as well as extensive testing post silicon with the processor in an actual platform \cite{bob}.
The algorithm described in this paper has been implemented in an Intel RIT generator,
used by verification
teams across multiple Intel architectures (Itanium, IA-32 and 64-bit IA-32). 
Although in the architecture validation (pre silicon on RTL simulators)
environment direct visibility into load and store execution allows simpler tools to be built, it has been used 
in a limited fashion to generate tests that
are subsequently run on RTL simulators. The results are then checked by the algorithm to find bugs. 
The greatest success of the tool has been in the Post Silicon Environment, 
where the execution speed available (compared to RTL simulations) allows the tool to 
quickly run a large number of random tests and discover memory ordering issues on processors.
In figure \ref{fig:example2} we show an example of an incorrect execution corresponding to
an actual bug found by this tool. The problem was subsequently traced to incorrect design in the CPU
of the locking primitive for certain corner cases.\\\\
In the Post Silicon environment the tool has been written to run directly on the Device Under Test(DUT).
This was made possible by running it as a process on a deviceless Linux kernel which is booted on the
target. The primary advantage of this model is speed and adaptability where the RIT tool directly detects
its underlying hardware, generates and executes the appropriate tests and then verifies the result 
with no communication overhead.Another not so apparent but important advantage is \emph{scaling}. 
As we anticipate future processors
to increase the number of available threads, the tool scales seamlessly by not only running tests on the 
increased number of threads but also using all available threads to run the checking algorithm itself.
This is also the reason why we have paid so much attention to parallelization in this work. That is to allow
the algorithm to bootstrap on future generations of multi threaded processors. We point out here that
the test generation phase is also parallelized in the tool to make optimal use of resources and achieve
the best speedup.\\\\
\textbf{Implementation Environment:}The algorithm is implemented in C 
and architecture dependent assembly that runs on a scaled down version 
of the Linux kernel. We have chosen to use
the Linux process model (avoiding other threading models for simplicity) with shared 
memory segments for inter process communication. We have hand parallelized the loops using the
data distribution concepts described in section \ref{sec:parallel}. This allows us to use
off the shelf compilers such as those in standard Linux distributions and work
across all the platforms that Linux supports.\\\\
\textbf{Exploiting SIMD:}
The key kernel used in the iterative phase of our algorithm is \\$Subsume$. This is called at least once
for every edge added to the graph and improving its performance is clearly beneficial. The implementation
for $Subsume(x,y)$ is\\ $\forall z \in \{1 .. n\} A[x,z]=A[x,z] \vee A[y,z]$. Another way of looking at it is
as the logical 'OR' of two binary vectors $A[x,.]=A[x,.] \vee  A[y,.]$. This could have taken as many as
$n$ operations in the most obvious implementation, but we instead chose to use Single Instruction Multiple
Data (SIMD) extensions available in both the IA-32 \cite{isa2} and Itanium \cite{itaniumisa} instruction sets.
These enable us to perform the subsume operation upto 128 bits at a time providing a 128 times
speedup to the implementation of $Subsume$. This is also the only place in our tool where we have
IA-32 and Itanium specific verification code.  The option to use SIMD to speedup the algorithm 
is really a consequence of the carefully selected data structures and the time consuming graph manipulations being 
reduced to a single well defined kernel.\\\\
\textbf{Extendibility:}
We support multiple architectures in our implementation by having as much architecture independent
code as possible. This means we need to only recompile the tool to target different architectures.
In addition we have made the tool independent of the memory consistency model it is verifying 
by taking as input to the tool a description of the local ordering rules, as described in 
definition \ref{def:local} in a standard format rulefile. This allows us to verify different 
consistency models (Itanium and different generations of IA-32) and adapt to changes in the 
consistency models that may happen in the future.\\\\
\textbf{Debug Support:}
A critical requirement in CPU verification is that failures should be root caused to 
bugs as soon as possible. Ease of debugging failures is very important in all of Intel's
verification methodologies. A failure in our case is a cycle in the graph. The problem 
with our algorithm formulation is that the final cycle is detected only in terms of 
\emph{which} nodes are participating in the cycle. There is no way to determine from the
closed form adjacency matrix what is the \emph{ordering} of nodes in the cycle. Also the
nature of the basic algorithm often leads to more than one cycle in long tests. To work 
around this problem without sacrificing algorithm efficiency we use a backtracking algorithm
described in Figure \ref{fig:debug} that prints all the detected cycles. The only change we 
need to make to the algorithm described in section \ref{sec:fa} is that it takes as parameter an edge $e$. 
Whenever the $AddEdge$ function adds the edge $e$ during execution of the algorithm we 
return true indicating that this edge is actually added by one of the rules in the algorithm. 
We also return the reason for addition of this edge which 
allows all edges to be labelled with the corresponding rule, a good aid to debug.
Note that the backtrack though costly is only run in case of failure which should be rare.

\section{Performance and Scaling}
\begin{figure}
\subfigure[Growth in cost with number of graph nodes]{
\includegraphics[height=4cm, width=6cm]{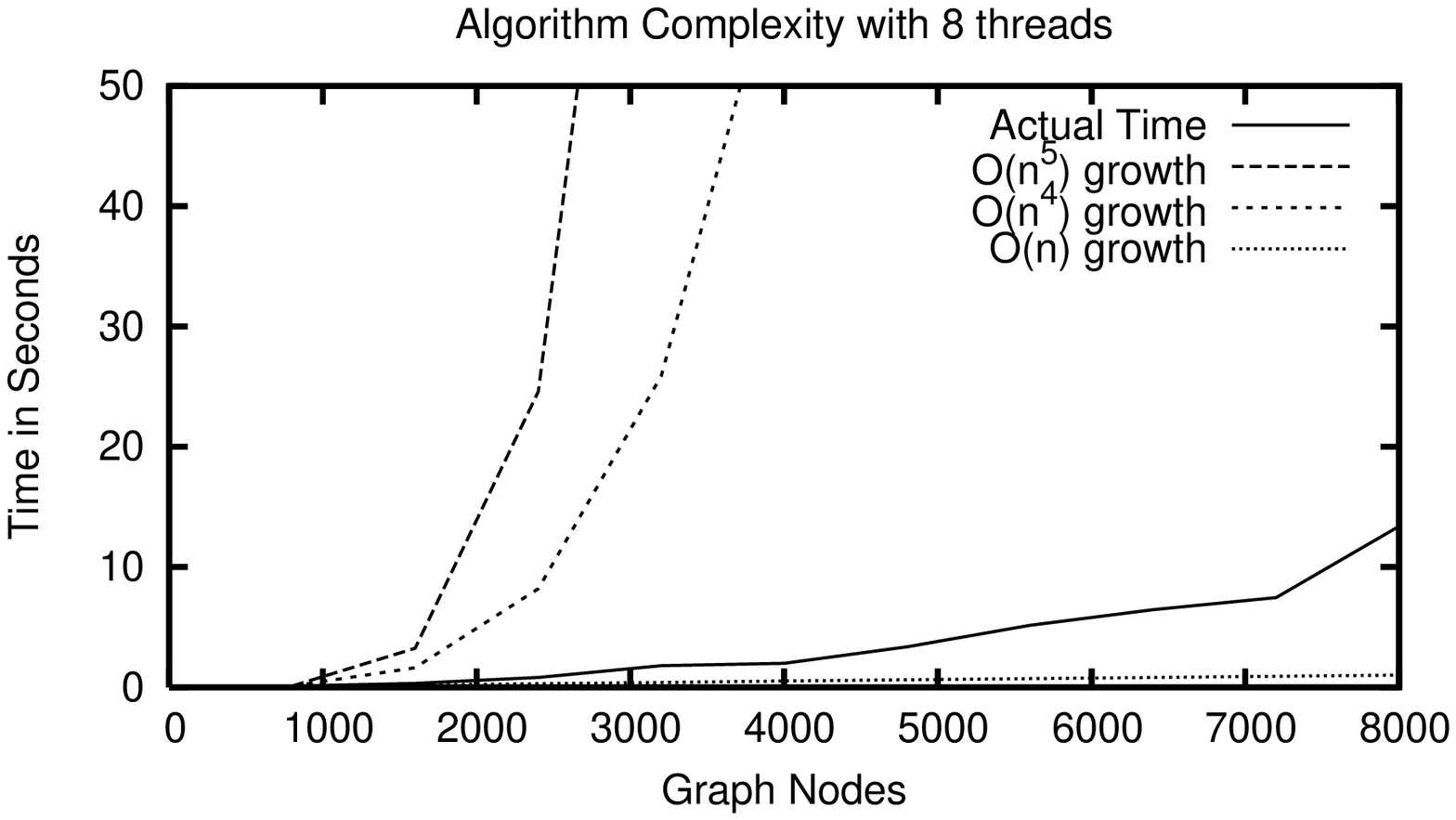}
\label{fig:complexity}
}
\subfigure[Speedup with increasing threads]{
\includegraphics[height=4cm,width=6cm]{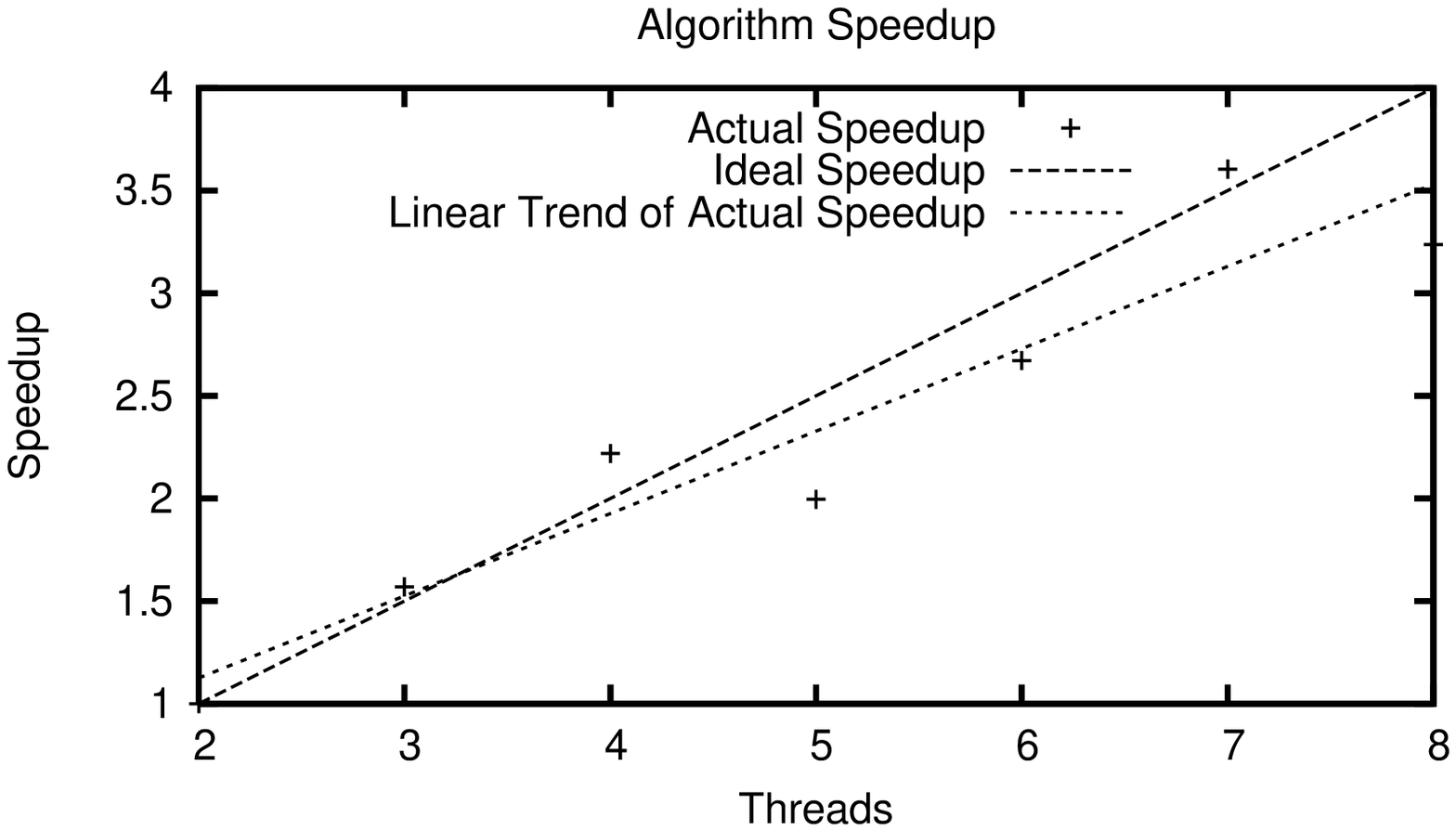}
\label{fig:speedup}
}
\caption{Algorithm Performance}
\end{figure}
We include some performance data to support our claims of efficient algorithm design.
In figure \ref{fig:complexity} we show how the cost of running the algorithm grows with increasing
number of nodes. Clearly the algorithm scales well. In figure \ref{fig:speedup} we show how
the speedup increases when we use more processors to run the algorithm while keeping the problem
size (number of graph nodes) same. The near to linear speedup (ideal) indicates that the parallelization
decisions have been correctly made and load balance the problem well among different processors.
All the presented scalability data was taken on an 8 way 1.2 Ghz Intel\Rtrademark Xeon\Rtrademark\footnote{Intel\Rtrademark Xeon\Rtrademark is a trademark or registered trademark of Intel Corporation or its subsidiaries in the United States and other countries.} processor platform running Linux.
\section{Limitations}
\begin{wrapfigure}{R}{5cm}
\subfigure{\framebox{\includegraphics[trim=6cm 6cm 6cm 6cm, clip, height=6cm, angle=270, width=4cm]{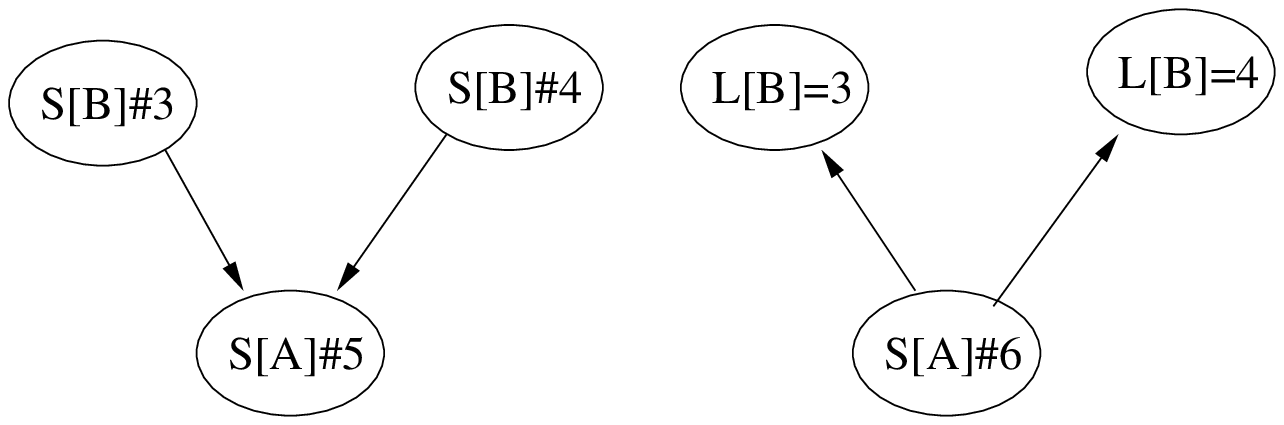}}}
\caption{Edge missed by the algorithm}
\label{fig:example3}
\end{wrapfigure}
Although our algorithm is general enough to cover the memory consistency models we need to check for
at Intel, it nevertheless has certain limitations and assumptions that we point out here.

We assume that all stores in the test to the same location write unique values. Thus we are never in a 
position where we need to reconcile a load with multiple stores for rule \ref{rl:obs}.

The algorithm assumes store atomicity, which is necessary for Axiom \ref{axm:tso}. However it supports
slightly relaxed consistency models which allow a load to observe a local store which precedes it in program order,
before it is globally observed. Thus we cover all coherence protocols that support the notion of relaxed write atomicity 
which can be defined as : No store is visible to any \emph{other} processor \emph{before} the execution point of the store. 
Based on our discussion with Intel microarchitects we determined that all IA-32 and \emph{current} generations of Itanium 
microprocessors support this due to identifiable and atomic global observation points for any store. This is mostly due to the 
shared bus and single chipset. For Itanium we can still adapt to the case where stores are not atomically observed by other
processors by checking only store releases \cite{itaniumspec}. Another approach is to split stores into one for each 
observing processor and appropriately modify rule \ref{rl:obs}. This would lead to a worse case degradation of checking 
performance by a factor equal to the number of processors. 

Last, the algorithm does approximate checking only (since it is a polynomial time solution to an NP-Hard problem).
It does not completely check for Axiom \ref{axm:tso}, since it does not attempt to order all stores and thereby
find additional inferred edges which could lead to a cycle. An example taken from \cite{tsotool} is shown in 
\ref{fig:example3}. The algorithm is unable to deduce the ordering from $S[A]\#6$ to $S[A]\#5$ although that is 
the only possibility given that the loads to location $B$ read different values. Adding a similar mirrored set 
of nodes, 2 stores to location C before $S[A]\#6$ and two loads from location C after $S[A]\#5$ give an example violation of the TSO model
which is missed by this algorithm. However, we hypothesize that only a small fraction of bugs actually lead to 
such cases and these are ultimately found by sufficient random testing which will show them up in a form the algorithm
can detect. This is well borne out in practice and another reason why we place so much emphasis on test tool performance.

\section{Conclusion}
We have described an algorithm that does efficient polynomial time memory consistency verification.
Our algorithm meets its stated goals of efficiency and generality. It is implemented in a tool that is
used across multiple groups in Intel to verify increasingly complex microprocessors. It has been appreciated
across the corporation for finding a number of bugs that are otherwise hard to find and point to extremely
subtle flaws in implementing the memory consistency model. We hope to work further in decreasing the cost of 
the algorithm by by studying the nature of the graphs generated
and considering more fine grained parallelization opportunities.\\
\textbf{Acknowledgments:} We would like to thank our colleagues Jeffrey Wilson and Sreenivasa Guttal for their
contribution to the tool, Mrinal Deo and Harish Kumar for their
assistance with memory consistency models and Hemanthkumar Sivaraj for giving valuable feedback 
during the initial stages of algorithm design.
\bibliography{paper}
\end{document}